\documentclass[%
mathleft,%
]{an}
\usepackage{graphicx}
\usepackage{times}
\begin{document}

\Pagespan{1}{}
\Yearpublication{2011}%
\Yearsubmission{2011}%
\Month{1}%
\Volume{999}%
\Issue{92}%

\title{Radio properties of Compact Steep Spectrum and GHz-Peaked Spectrum radio sources}

\author{M. Orienti\inst{1}\fnmsep\thanks{Corresponding author:
  \email{orienti@ira.inaf.it}}
}
\titlerunning{Radio properties of CSS and GPS radio sources}
\authorrunning{M. Orienti}
\institute{
INAF-IRA, Via Gobetti 101, 40129 Bologna, Italy
}
\received{XXXX}
\accepted{XXXX}
\publonline{XXXX}


\abstract{%
  Compact steep spectrum (CSS) and GHz-peaked spectrum (GPS) radio
  sources represent a large fraction of the extragalactic objects in
  flux density-limited samples. They are compact, powerful radio
  sources whose synchrotron peak frequency ranges between a few
  hundred MHz to several GHz. CSS and GPS radio sources are currently
  interpreted as 
  objects in which the radio emission is in an early evolutionary
  stage. In this contribution I review the radio properties and the
  physical characteristics of this
  class of radio sources, and the interplay between their radio emission and the
  ambient medium of the host galaxy. } 

\maketitle

\section{Introduction}

Compact steep spectrum (CSS) and GHz-peaked spectrum (GPS) radio
sources are powerful (P$_{\rm 1.4\ GHz}>$10$^{25}$ W/Hz) and compact
objects with angular sizes not exceeding 1 -- 2 arcsec. 
The main peculiarity of these
objects is the convex synchrotron radio spectrum that peaks
around 100 MHz in the case of CSS sources, and at about 1 GHz in the
case of GPS objects, or even up to a few GHz in the sub-population of
high frequency peakers (HFP) defined by Dallacasa et al. (2000). 
Above the peak frequency the spectrum is steep
with a spectral index $\alpha \sim
0.7$ ($S_{\nu} \propto \nu^{- \alpha}$). \\
Depending on both the frequency and the flux-density limit
of the catalogues used, the CSS/GPS samples are dominated by different
sub-classes of objects. Bright CSS and GPS samples have been selected
from the 3C, PW and 1-Jansky catalogues (see, e.g., Spencer et al. 1989,
Fanti et al. 1990, Stanghellini et al. 1998). On the other hand, deep
catalogues, like B3, FIRST, WENSS and AT20GH, were used for selecting weak
samples (e.g., Fanti et al. 2001, Snellen et al. 1998, Kunert et
al. 2002, Hancock et al. 2010). In the last decades other samples
of CSS/GPS candidates were constructed using different selections tools like the radio
morphology, optical counterpart, compact linear sizes (e.g., COINS sample, Peck \&
Taylor 2000; CSS-VIPS sample, 
Tremblay et al. 2009; CORALZ, Snellen et al. 2004), as well as
polarization properties (Cassaro, Dallacasa \& Stan\-ghellini 2009).\\
Statistical analysis of CSS/GPS samples pointed out an empirical
anti-correlation between the peak frequency of the spectrum and the
linear size (O'Dea \& Baum 1997). This has been interpreted either in terms
of synchrotron-self absorption related to the compact dimension of the
sources (e.g., Snellen et al. 2000, Fanti 2009, Orienti \& Dallacasa
2008a) or due  
to free-free absorption 
from an ionized medium enshrouding the radio emission (e.g., Bicknell
et al. 1997, Tingay et al. 2015, Callingham et al. 2015), although a
combination of both 
mechanisms may take place 
(e.g., Orienti \& Dallacasa 2008a).\\ 
CSS and GPS radio sources represent a significant fraction (15\% -- 30\%
depending on the frequency) of the sources in flux-density limited
catalogues, opening a debate about their nature. 
Fanti et al. (1990) investigated
whether the compact size is a result of projection effects, but they
concluded that this was unlikely, leaving room only for a minority
($<$25\%) of large objects foreshortened by geometrical effects.
The intrinsically compact size of CSS/GPS is interpreted mainly in
terms of {\it Youth}: these sources are small because they are still
in an early stage of their evolutionary path, and may become/develop
into Fanaroff-Riley type-I/II (FRI/FRII, Fanaroff \& Riley 1974) radio
sources (e.g., Fanti et al. 1995, Snellen et 
al. 2000, Alexander 2000, Perucho 2015). Strong supports to the {\it
  Youth Scenario} came from the estimate of the kinematic
age by the determination of the hot spot separation velocity in a
handful of the most compact objects (Polatidis 2009, Giroletti \&
Polatidis 2009, and references therein, Polatidis \& Conway 2003, and
references therein), as well as from the radiative age (Murgia 2003,
Murgia et al. 1999, Nagai et al. 2006).
The alternative scenario that postulated the presence of an
exceptionally dense medium able to frustrate the jet growth was not
supported by multiband observations which pointed out that the gas of
their host galaxies are similar to those of extended FRII sources
(e.g., Fanti et al. 1995, 
Fanti et al. 2000, Siemiginowska et al. 2005).\\

In the next Sections I will briefly review the observational and
physical properties of CSS and GPS radio sources. Radio properties 
are presented in
Sections 2. In Section 3 I describe the physical parameters, such as the
luminosity and the magnetic field, and how they evolve.
In Section 4 I discuss the duty cycle of the
radio emission, while in Section 5 and 6 I present the characteristics
of the ambient medium and their role in producing the observed source
asymmetries. A brief Summary is
presented in Section 7.\\
Throughout the paper I assume H$_0$ = 71 km s$^{-1}$ Mpc$^{-1}$,
$\Omega_{\rm M}$ = 0.27, $\Omega_{\Lambda}$ = 0.73 in a flat
Universe. The spectral index is defined as $S_{\nu} \propto \nu^{- \alpha}$. \\

\section{Radio properties}
\label{properties}

\subsection{Morphology}

\begin{figure*}
\begin{center}
\includegraphics{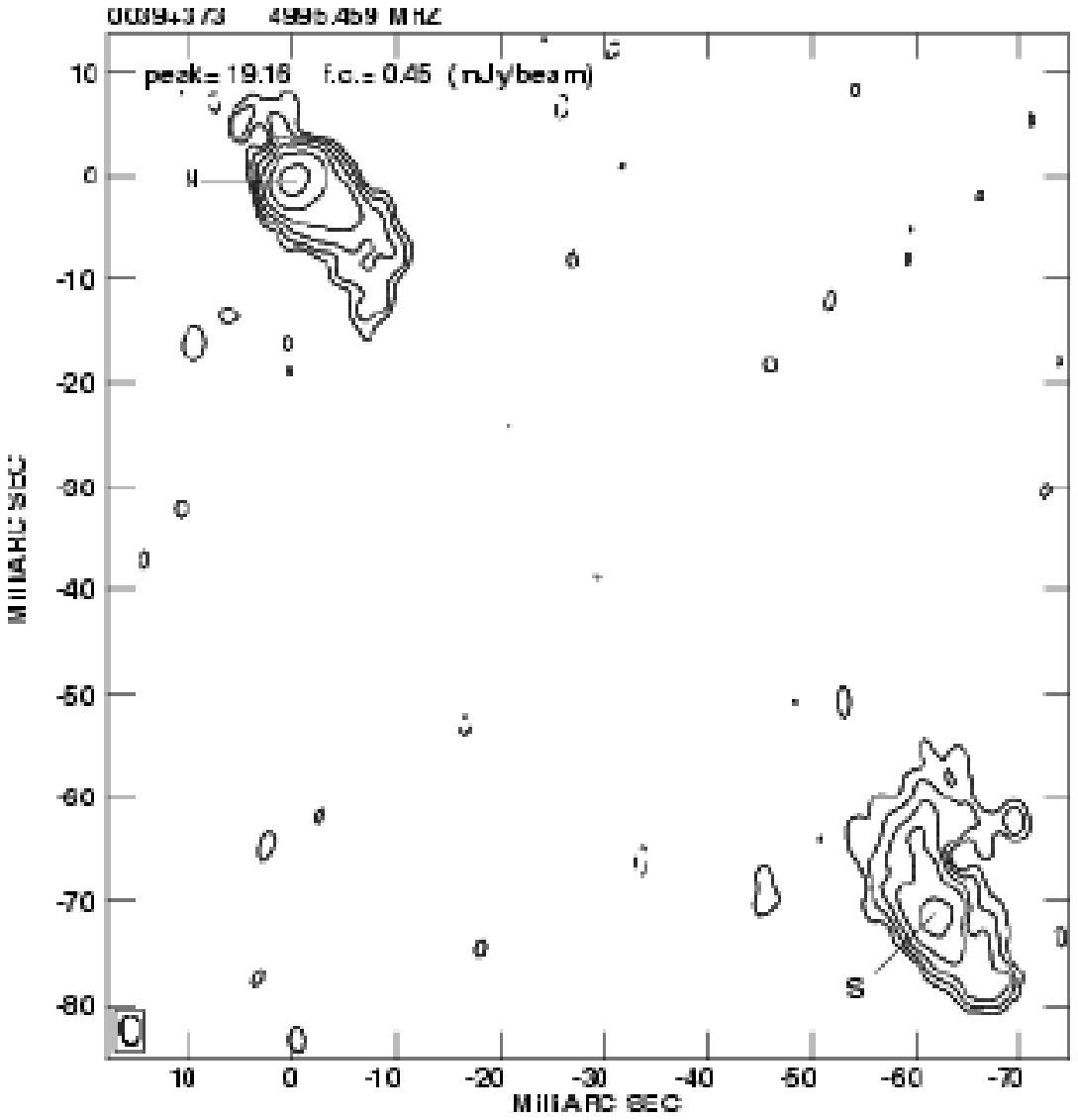}
\includegraphics{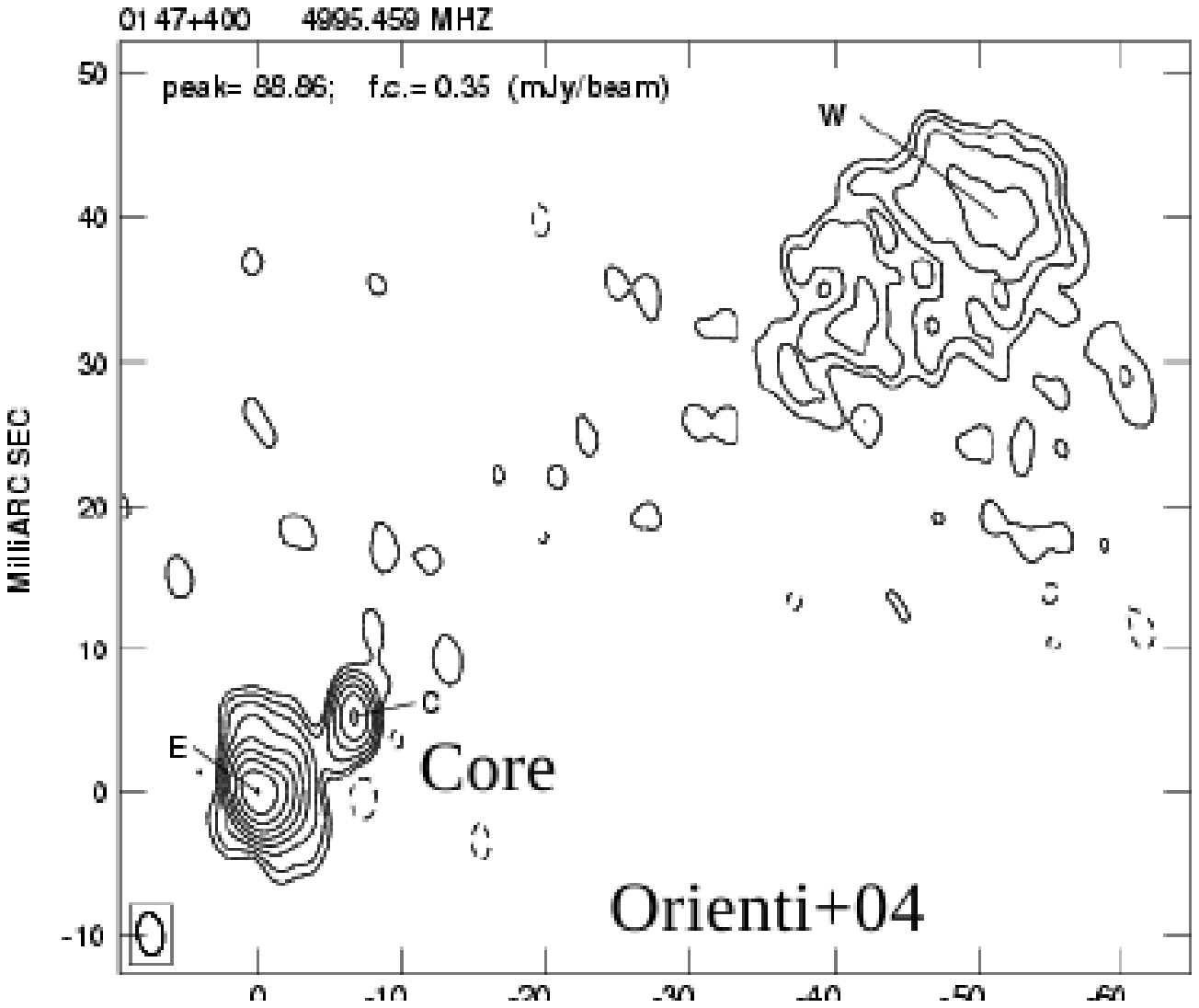}
\vspace{6cm}
\caption{Examples of a symmetric ``two-sided'' CSO ({\it left}), and
  of an asymmetric one ({\it right}). Adapted from Orienti et
  al. 2004}
\label{morpho}
\end{center}
\end{figure*}

Due to their compact size, CSS/GPS sources appeared unresolved in
single-dish observations, and only the advent of interferometers with
sub-arcsecond resolution could pinpoint their radio morphology. 
CSS and GPS are divided into three main morphological classes: 1)
symmetric (i.e. two-sided) structures; 2) core-jet structures; 3)
complex morphology.\\
Symmetric objects have a two-sided radio structure resembling
a scaled-down FRII radio source (Fig. \ref{morpho}). 
The main ingredients are mini-lobes
and hot spots. Sometimes, a weak component hosting the core is
present, and, depending on their size, CSS/GPS may be
termed as ``compact symmetric objects'' (CSO) if they are smaller
than 1 kpc, 
or ``medium-sized symmetric objects'' (MSO) if they extend up to 10 --
15 kpc (Fanti et al. 2001). 
However, a large fraction of ``symmetric'' sources have a very
asymmetric two-sided morphology, where one side of the source is much
brighter than the other (e.g. Saikia et al. 2003, Rossetti et
al. 2006). When the core is 
detected it is usual to find that the brighter lobe is the one closer
to the core. This is opposite to what is expected in presence of
geometrical effects, suggesting some interaction between the jet and
the ambient medium (see Section \ref{asymmetry}). \\ 
In two-sided objects the core
usually represent a very small fraction (a few per cent) of the total
radio emission of the source indicating the absence of beaming effects
(e.g. Wilkinson et al. 1994).
On the contrary, in radio sources with a core-jet and complex structure the core
dominates the radio emission, suggesting the presence of significant
boosting effects. In addition core-jet structures are usually found in
CSS/GPS which are optically identified with quasars (e.g., Rossetti et
al. 2004, Orienti et al. 2006), supporting the role of projection effects. \\
As in the case of core-jet structures, also complex morphologies are
usually caused by boosting effects, at least in high-luminosity
sample. This does not seem to hold in low-luminosity samples where a high
fraction of sources ($\sim$30\%) have weak 
extended emission and distorted structures which are likely intrinsic
(Kunert-Bajraszewska et al. 2010). \\

\subsection{Variability}

The spectral behaviour and the flux density variability is
investigated by multifrequency observations, closely simultaneous when
possible, carried out at various epochs. In past works CSS/GPS objects
were
considered the least variable extragalactic radio sources (O'Dea
1998). However, long-term monitoring campaigns found different results
depending on the different sub-classes considered (CSS, GPS, HFP, quasars,
galaxies).
It turned out that many GPS/HFP sources are picked up with a convex
spectrum only during flaring events, when the radio emission is
dominated by the jet, while in their average state they possess a flat
spectrum (Torniainen et al. 2005, Tinti et al. 2007).
In particular, high fraction of CSS radio galaxies are not variable, while
only $\sim$30\% of GPS/HFP galaxies preserve the convex spectrum
(Torniainen et al. 2007, Orienti et al. 2010, Hancock et al. 2010). 
The majority of the CSS, GPS, and HFP quasars show
significant flux density and spectral variability 
(Mingaliev et al. 2012, Orienti et
al. 2007). \\
Spectral changing and flux density variability do not always imply
that the source is part of the blazar population, rather than a
genuine CSS/GPS/HFP object. In fact, changes in the radio spectrum may
be a direct consequence of the source expansion (e.g., Tingay \& de
Kool 2003). In newly born radio sources, the evolution time-scales can
be of the order of a few tens of years. Changes in the radio spectrum
of such young objects can be appreciable after the short time (5 -- 10
yr) elapsing between the observing epoch. If the variability  is due
to the source expansion we expect that the peak
shifts to lower frequencies, the flux density in the optically-thin
regime decreases, while that in the optically-thick part of the
spectrum increases. This behaviour has been observed in the HFP
RXJ1459+3337 (Orienti \& Dallacasa 2008b), as well as in a handful of
HFP/GPS sources (Dallacasa \& 
Orienti 2015, Orienti et al. 2010), although some additional
variations in the free-free optical depth may be present (e.g., the GPS
PKS 1718-649, Tingay et al. 2015). \\

\subsection{Polarization}

CSS/GPS objects are weakly polarized. Multifrequency
polarimetric measurements of a sample of CSS/GPS radio sour\-ces (Fanti
et al. 2004, Cotton et al. 2003) show that very compact objects ($<$1
kpc) are unpolarized or strongly depolarized, and the fractional
polarization is strictly related to the frequency: the lower the
frequency, the stronger the depolarization. Furthermore, the
fractional polarization does not increase gradually with the source
size, but there seems to be a discontinuity at a ``critical size''
(the so-called {\it Cotton effect}),
that is about 6 kpc at
1.4 GHz and moves down to about 1 kpc at higher frequencies. \\
The strong
depolarization of the most compact objects may be related to the
interstellar medium of the narrow line region (NLR) which acts as a
Faraday Screen, depolarizing and/or rotating the polarized signal. The
``amount'' of depolarization and/or rotation depends on both the
inhomogeneities of the ambient medium and the distribution of the
magnetic field. Support to the presence of a dense and inhomogeneous
medium in front of the radio source comes from the large rotation
measure (RM) estimated for those sources with some polarized
emission. RM of the order of 1000 rad m$^{-2}$ or even higher are
commonly observed (e.g., O'Dea 1998, Cotton et al. 2006, Rossetti et
al. 2008, Mantovani et al. 2013).\\
A different result was found by Mantovani et al. (2009) who studied a
complete sample of CSS with multifrequency single-dish
observations. In this case no drop in the polarization was found for
the most compact sources. However, this apparent contradiction is
likely due to the contamination by geometrically-foreshortened
quasars. In fact, if the galaxies are considered separately, the {\it
  Cotton effect} is visible again, while the fractional polarization
in quasars seems independent from the linear size. In addition, quasars
have higher fractional polarization than galaxies, while galaxies
experience larger RM. In HFP quasars the high
fractional polarization is associated with low RM and high flux
density variability (Orienti \& Dallacasa 2008c), similar to what is
found in blazars (Mantovani, Bondi, \& Mack 2011). 
Sub-arcsecond resolution
observations point out higher RM than those estimated by
low-resolution single-dish observations suggesting the presence of
blended components. \\
Another intriguing aspects pointed out by multifrequency observations
is an increase of the polarized emission at low frequency
(re-polarization). This may be explained either in terms of multiple
unresolved components with different spectral characteristics that
dominate at different frequencies, or in presence of variability
(Mantovani et al. 2009, Orienti \& Dallacasa 2008c). In this case the
RM does not follow the $\lambda^{2}$ correlation (see e.g., Burns 1966 ).\\
A remarkable result found is that CSS are more asymmetric in the
polarization of the outer
lobes than the extended galaxies (Saikia \& Gupta 2003). 
The high incidence of polarization
and morphological asymmetries observed in CSS and GPS is likely an
indication of jet-gas interaction which is more probable when the
radio jet is still piercing its way through the dense medium of the
host galaxy (see Section \ref{asymmetry}).\\

\section{Physical properties}
\label{physical}

In a scenario where radio sources grow in a self-similar way, the
evolution of each radio object originated by an AGN depends on its
linear size. 
The determination of the physical properties in objects at the
beginning of their evolution is crucial for setting tight constraints
on the initial conditions of the radio emission. \\
The source size, flux density, and the peak frequency are parameters
that can be easily derived from the observations and then used as
a starting point to determine the physical properties of the
sources. \\
The existence of a 
relation between the rest-frame peak frequency and the projected
linear size (e.g. O'Dea \& Baum 1997) indicates that the mechanism responsible
for the curvature of the spectrum is related to the source dimension,
and thus to the source age. Interestingly, some of the most compact
and asymmetric sources seem to depart from this relation. However, this
is likely due 
to the presence of several sub-components that are responsible for different
part of the total radio spectrum: one bright and compact hot spot
dominates the radio emission, overwhelming the contribution from the
extended structures (e.g. J1335+5844, Orienti \& Dallacasa 2014). \\
Another important relationship to investigate is between the peak flux
density and the peak frequency, since they provide constraints on the
magnetic field in case the spectral curvature is due to SSA. 
The analysis of the peak flux density as a
function of the peak frequency in CSS/GPS/HFP sources from bright
samples suggests a segregation between sources identified with galaxies
and quasars. When only galaxies are considered
there seems to be an anticorrelation between the peak flux density and
the peak frequency, as expected if the spectral turnover is due to
SSA, although FFA cannot be completely discarded. On the contrary,
this anticorrelation does not hold in case of 
quasars. They have peak frequency in GHz regime, while the
peak flux density covers three order of magnitudes, independently of
the peak frequency, suggesting a different nature for the majority of
galaxies and quasars (Orienti et al. 2010). \\

\subsection{Magnetic field}

The direct measurement of the magnetic field in extragalactic radio
sources is a difficult task to carry out. An indirect way to estimate
the magnetic field is to assume that the radio source is in
minimum energy condition corresponding to a near equipartition of
energy between the 
radiating particles and the magnetic field (Pacholczyk 1970). 
Although this condition is assumed in many evolutionary models,
there is no a priori reason for believing that magnetic fields in
radio sources are in equipartition.  \\
As mentioned earlier, a direct measurement of the magnetic field from
observable quantities is obtained by means of the spectral parameters. If the
spectral peak is produced by SSA, we can compute the magnetic 
field $H$ by using observable quantities only:

\begin{equation}  
H \sim f(\alpha)^{-5} \theta^{4} S_{\rm p}^{-2} \nu_{\rm p}^{5} (1+z)^{-1}
\label{magnetic}
\end{equation}  

\noindent where $\theta$ is the source solid angle, $\nu_{\rm p}$ and
$S_{\rm p}$ are the peak frequency and peak flux density,
respectively, $z$ is the redshift, and $f$($\alpha$) is a function
that depends weakly on the spectral index (Kellermann \& Pauliny-Toth
1981). 
Scott \& Readhead (1977) and Readhead (1994) computed the magnetic
field for sources of low-frequency spectral turnovers close in value
to the observing frequency and found that the magnetic fields inferred
directly from the spectrum were within a factor of 16 of the
equipartition values. However, there are no systematic studies of
sources with spectra peaking at higher frequencies, i.e. objects
younger than those in Scott \& Readhead (1977). \\ 
The main difficulty in applying this method has been the
uncertainty in determining source component parameters at the turnover
frequency, which results in a limited accuracy of the magnetic field
estimates. However this method may be used for GPS/HFP sources. The
peak frequency around a few GHz gives the possibility to sample both
the optically-thick and -thin part of the spectrum by multifrequency
high resolution VLBA observations, leading to a fairly accurate
estimate of the peak parameters.\\
The peak magnetic fields estimated for the components of HFP radio
sources turned out to 
be in good agreement with those derived by assuming minimum energy
conditions, supporting the idea that in general young radio sources
are in minimum energy conditions and their spectral turnover is caused
by SSA (Orienti \& Dallacasa 2008a, 2008b, 2014). However, there are a
few exceptions where the peak magnetic field is orders of
magnitude higher than the equipartition value. The analysis of the
optically-thick part of the spectrum turned out to be more inverted
than the limit value for SSA ($\alpha < - 2.5$), indicating that the
spectral peak is due to FFA. Therefore the magnetic field determined
by the peak values is meaningless. \\ 
Depending on the size of the sources, the estimated magnetic fields
range from a few 150 mG in the most compact components of HFP radio
sources down to 0.1 mG in the lobes of CSS (Fanti et al. 2001,
Dallacasa et al. 2002, Orienti et al. 2006). The anticorrelation
between the magnetic field strength and the linear size is in
agreement with what is expected 
in case the source is adiabatically expanding. However, when the
single source components are considered, it emerges that the field
intensities found in the various components of the same object can
vary up to an order of magnitude (Orienti \& Dallacasa 2012). Such differences may arise from
asymmetries in the source propagation, for example when
the two sides 
experience a different environment (see Section \ref{ambient}),
indicating that simple 
self-similar evolution models may be not adequate to describe the
radio source growth (e.g., Sutherland \& Bicknell 2007).\\

\subsection{Luminosity evolution}  

\begin{figure}
\begin{center}
\includegraphics{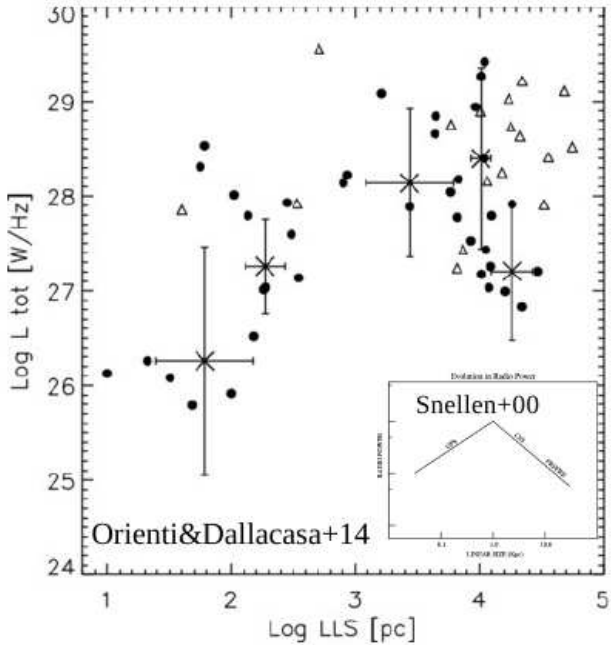}
\vspace{9cm}
\caption{Total luminosity at 375 MHz versus the linear size for the
  sources of the sample selected by Orienti \& Dallacasa
  (2014). Filled circles are galaxies, while empty triangles are
  quasars. Crosses represent the median values of the total luminosity
  and linear size, for galaxies only, separated into different bins. As a
  comparison, in the bottom right corner there is the evolutionary
  trend expected by the model developed in Snellen et al. (2000).} 
\label{lum_ls}
\end{center}
\end{figure}

Several evolutionary 
models\footnote{The evolutionary models considered in this review try
  to explain how an  
individual radio source grows and evolves, without taking into account any cosmological implication.} (e.g. Fanti et al. 1995; Readhead et al. 1996; Snellen et
al. 2000) were proposed to describe  
how the physical parameters (i.e. luminosity, linear size and
velocity) evolve as the radio emission grows within the host galaxy. 
The majority of the proposed models predict an increase of the
luminosity and a decrease of the jet advance speed when the radio
emission is still embedded within the dense medium of the NLR. Then,
as the radio emission emerges from the NLR the luminosity is expected
to decrease, while the jet advance speed should not vary
significantly. These predictions should be validated by statistical studies of
samples of CSS/GPS/HFP spanning a large 
range of linear size.\\
O'Dea (1998) studied the radio luminosity at 5 GHz as a function of
the projected linear size, but no correlation was found. This may be
related to the similar radio power of the selected sources.
To determine how the luminosity evolves as the source grows, Orienti
\& Dallacasa (2014) studied a sample of 51 {\it bona fide} young radio
sources with an unambiguous detection of the core region, and spanning
a wide range of linear size, from a few pc up to tens of kpc, for
high-luminosity radio sources. To get rid of possible
boosting/projection effects 
that may contaminate the estimate of the physical parameters, only
objects optically associated with galaxies were considered when
searching for empirical relations.\\
The analysis of the source luminosity at 375 MHz versus the linear size points
out two different relations depending on the source size: sources
smaller than a few kpc increase their luminosity as they grow, while
larger sources progressively decrease their luminosity, in agreement
with the evolutionary models (Fig. \ref{lum_ls}). 
The smallest sources reside within the
innermost region of the host galaxy, where the dense and inhomogeneous
ambient medium favours radiative losses. As the radio source expands
on a kpc scale, it experiences a smoother and less dense ambient
medium and adiabatic losses dominate. \\
Kunert-Bajraszewska et al. (2010) extended the analysis of the
radio power as a function of linear size by including a sample of
low-luminosity CSS objects having 1.4-GHz luminosity comparable to
that of FRI. They found a clear distinction between high-power and
low-power objects: the former seem to follow an evolutionary path
similar to that expected by the models. The latter are located below
this path and they may represent short-lived objects with a different
fate. \\

\section{The life-cycle of the radio emission}
\label{duty}

\begin{figure*}
\begin{center}
\includegraphics{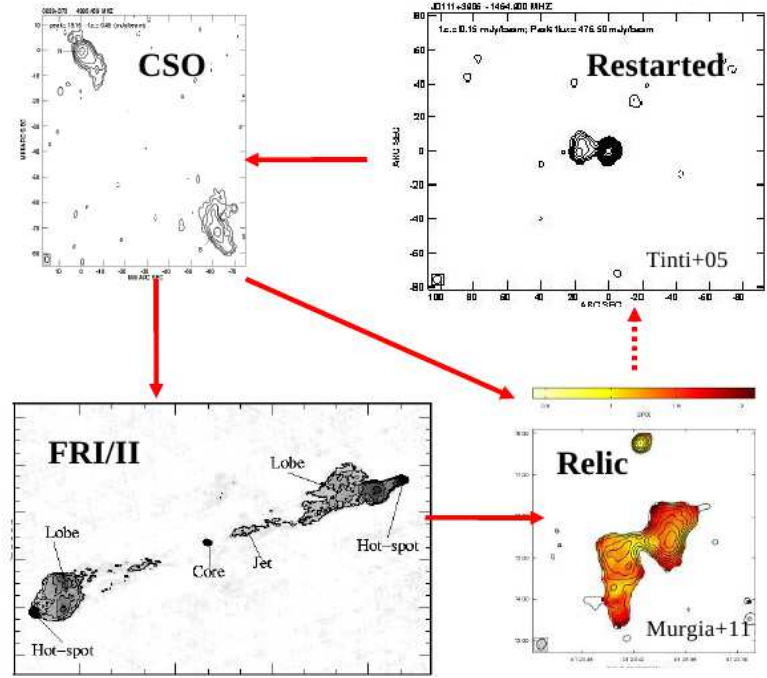}
\vspace{9.5cm}
\caption{The evolutionary path of the radio emission. Young CSO ({\it
    top left}, image adapted from Orienti et al. 2004) may become
either a classical large FRII ({\it bottom 
    left}, image adapted from Mack et al. 2009) or a relic in the case
  of the activity phase switches off 
  soon after its onset ({\it bottom right}, image adapted from Murgia
  et al. 2011). If the central engine
  goes through another active phase, a newly-born bright and compact
  object can be observed close to the relic of the previous activity
  ({\it top right}, image adapted from Tinti et al. 2005).}
\label{evolution}
\end{center}
\end{figure*}

It is nowadays clear that powerful (L$_{\rm 1.4~GHz} >
10^{25}$ W/Hz) radio sources are a small fraction of the
AGN generally associated with ellipticals, suggesting that the radio activity is
a transient phase in the life of these systems. The typical age of
active phase in radio sources is about 10$^{7}$ -- 10$^{8}$ years,
which is followed by a relic phase which is roughly one order of
magnitude shorter (Parma et al. 2007).\\ 
The onset of radio emission is currently thought to be
related to mergers which provide fuel to the central
AGN. However, the reason why and when the radio emission swit\-ches off
is still an open question. The excess of young objects in flux-limited
samples suggests the existence of short-lived objects
unable to become FRII, and additional ingredients,
like the recurrence of the radio emission (e.g., Czerny et al. 2009), or the
interplay between the source and the environment, 
must be considered (see Section \ref{ambient}). 
Support to the existence of short-lived objects
come from statistical studies of the ages of CSO which peak at about
500 years (Gugliucci et al. 2005), and of the sub-class of
low-luminosity CSS radio sources (Kunert-Bajraszewska et al. 2010).\\
If the supply of new relativistic particles turns
off, the radio emission fades rapidly due to the severe energy
losses and the radio spectrum steepens fast making these sources
under-represented in flux-limited catalogues. Indeed, only a few
objects have been suggested as faders so far, based on the absence 
of active regions (Kunert-Bajraszewska et al. 2005, 2006), and the
distribution of spectral index found steep across the whole source, 
like in the case of PKS \\ 1518+047 (Orienti, Murgia, \& Dallacasa 2010).\\
A different situation is the recurrence of the radio emission in an
AGN. In this case a ``young'' radio source, with new activity regions
like the core and hot spots, is present close to the fossil of a
previous epoch of the radio emission. A clear example of intermittent
radio activity is the FRII radio galaxy B0925+420 where three
different episodes of jet activity have been observed (Brocksopp et
al. 2007).\\
Extended emission on the kpc-scale and beyond was discovered in the
GPS galaxy J0111+3906 (Baum 1990), and interpreted in terms of the
relic of a past radio activity which occurred about 10$^{7}$-10$^{8}$ years ago.
Recently, a remnant of about 160 kpc in size from an earlier stage of
activity was found in the 
CSS galaxy B2\,0258+35, suggesting that the time between subsequent
phases of activity in this source is about 10$^{8}$ (Shulevski et
al. 2012, Brienza et al. 2015).\\
Following the model by Czerny et al. (2009), the restarting of the
radio emission may occur on much shorter (a few 
thousand years) time scales. This is the case of the HFP
sources J1511+0518 and OQ\,208 where a relic from the past activity is
found at about 50 pc from the reborn object (Orienti \& Dallacasa
2008, Luo et al. 2007), indicating that, at least, in some objects the
duty-cycle of the radio emission occurs on time-scale of 10$^{3}$ --
10$^{4}$ years.\\
Following the evolutionary path, CSO should evolve into FRII radio
galaxies with ages of 10$^{7}$ --10$^{8}$ years before the radio emission
enters in the relic phase. However, it is
possible that not all the CSO would become FRII, and a population of
fading short-lived objects under-represented in flux-limited catalogue
is expected. If the interruption of the radio activity is a temporary
phase and the radio emission from the central engine will restart
soon, it is possible that the source will appear again as a CSO
without the severe steepening at high frequencies. If this does not
happen, the fate of the fading radio source is to emit at lower and lower
frequencies, until it disappears at frequencies well below the MHz
regime (Fig. \ref{evolution}).  \\

\section{Ambient medium}
\label{ambient}

CSS and GPS radio sources are usually found in early-type gas-rich
galaxies. The presence of significant amount of gas in young radio
galaxies is supported by a larger incidence of HI absorption in these
objects (Vermeulen et al. 2003, Pihlstr\"om, Conway, \& Vermeulen 2003, Gupta et
al. 2006) compared to
what is typically found in old and larger radio sources (Morganti et
al. 2001).
This dense medium is likely the result of the merger that
triggered the radio source (Morganti et al. 2004a). The knowledge of
the distribution of the gas, either settled in a circumnuclear structure like a disk or
a torus, or inhomogeneously distributed in clouds, provides important
information on the environment in the innermost region
of AGN and its role in the radio jet evolution. \\
Pihlstr\"om et al. (2003) found an empirical anti-correlation between
the HI column density, $N_{\rm HI}$ and the source linear size in a sample of 41
CSS/GPS sources: smaller sources ($<$0.5 kpc) have larger HI column
densities than the larger sources ($>$0.5 kpc). This result suggests
that the HI gas is likely settled in a circumnuclear disk/torus with a
radially declining density, and
the absorption takes place only 
against the receding lobe. It is worth noting that the $N_{\rm
  HI}$-$LS$ anticorrelation does not seem to hold in the most compact
HFP objects (Orienti, Morganti, \& Dallacasa 2006), which highly
deviate from the trend by Pihlstr\"om et 
al. (2003). The absence of high HI column density in very compact
sources can be explained by both the orientation and the extreme
compactness of the sources in a disk/torus scenario, in which our line
of sight intersects the circumnuclear structure in its inner region
where the low optical depth may be due to high spin and kinetic
temperatures.  \\
The presence of a gas distribution consistent with a circumnuclear
disk/torus is supported by high-resolution VLBI observations that
could trace either the atomic hydrogen across the source (e.g., 4C\,31.04,
Conway 1999; PKS\,1946+708, Peak \& Taylor 1999; B2352+495, Araya et
al. 2010) 
or the ionized gas via free-free absorption (J1324+4048 and
J0029+3457, Marr et al. 2014; J0111+3906, Marr, Taylor, \& Crawford
2001), and by high-sensitivity observations of key molecular species
such as CO, HCN, and HCO$^{+}$ (e.g., 4C\,31.04, Garcia-Burillo et
al. 2009; OQ\,208, Oca\~na-Flaquer et al. 2010).\\
Not all the ionized, atomic, and molecular gas is organized in a
circumnuclear structure. As a consequence of the recent
merger/accretion events experienced by the host galaxy, we expect the
presence of unsettled, clumpy gas inhomogeneou\-sly distributed
particularly in the innermost region. However, a secure prove of such
clouds is difficult to obtain usually due to their low optical depth and
small filling factor. However, if the cloud impacts with the jet 
temporarily confining its expansion. Shallow and broad HI absorption
lines suggesting a jet-cloud interaction, were found in young or
restarted objects, but the lack of angular resolution prevented a
reliable interpretation of their origin, either circumnuclear or
off-nuclear (Morganti, Tadhunter \& Oosterloo 2005). 
An outstanding example is
represented by the young, restarted GPS
4C\,12.50, where VLBI observations locate the HI outflow of about
10$^{4} M_{\odot}$ at 
about 100 pc from the 
nucleus where the radio jet interacts with the ISM, as well as around
the associated radio lobe (Morganti et al. 2013). A possible molecular
counterpart of the atomic outflow was found by Dasyra \& Combes (2012)
by deep CO observations. Interestingly, in
the same source HI is observed in absorption also against the other
lobe and may cause the bending of the jet (Morganti et
al. 2004b). These results indicate that, despite the small filling
factor of the clumpy medium, jet-cloud interaction can take place and
may be responsible for some asymmetric radio sources observed. This
may be the case of the two highly asymmetric CSS sources 3C\,49 and 3C\,268.3,
in which HI absorption is observed only against the brightest lobe,
which is also the closest to the core, although the non-detection in
the other lobe might be due to sensitivity limitation (Labiano et
al. 2006). \\

\section{Asymmetries}
\label{asymmetry}

\begin{figure}
\begin{center}
\includegraphics{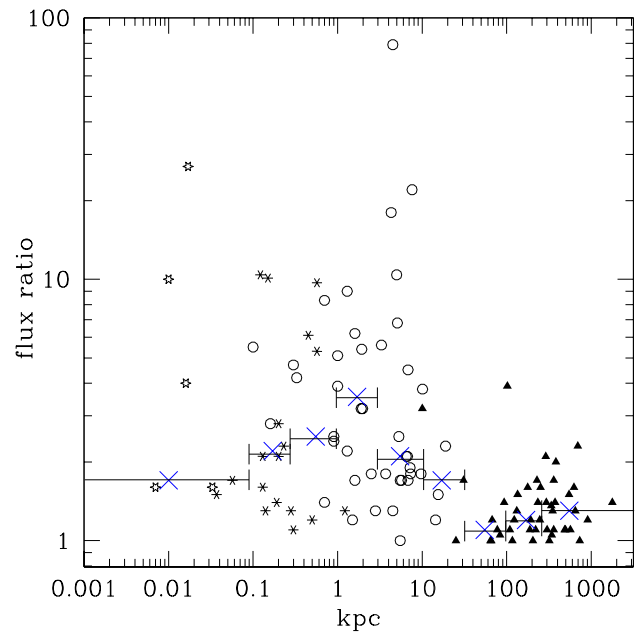}
\vspace{8.5cm}
\caption{The flux density ratio of the lobes versus the source linear
size. Crosses represent the median value computed on various
range of linear sizes. Adapted from Orienti \& Dallacasa (2008d).}
\label{asymmetries}
\end{center}
\end{figure}

As mentioned in Section \ref{properties}, a large fraction of CSS/GPS
sources have a very asymmetric structures (e.g., Saikia et
al. 2003). A significant fraction 
($\sim$50\%) of
asymmetric sources have the brighter lobe that is also the closer to
the core, which is opposite to what is expected if the source is not
on the plane of the sky and some beaming effects and path delay are
present. The brighter-when-closer trend  
 suggests that the two jets are piercing their way through an
 inhomogeneous medium as pointed out by HI observations (see Section
 \ref{ambient}). The interaction between the advancing jet and a
 clumpy medium may enhance the luminosity due to high radiative losses
 which become predominant with respect to the adiabatic
 ones. Jet-cloud interactions should be more frequent during the first
 stages of the radio emission, when the jet is piercing its way
 through the dense and inhomogeneous gas of the host
 galaxy. Fig. \ref{asymmetries} shows the flux density ratio versus the
 linear size of a sample of CSS/GPS and FRII (Orienti \& Dallacasa
 2008d). Sources larger than $\sim$15 kpc are more symmetric than
 the smaller ones.\\ 
The enhancement of the flux density may explain part of the high
number counts of CSS/GPS objects in flux density limited
samples. Furthermore, although the jet-medium interaction may not
frustrate the source expansion for its whole lifetime, it may severely
slow down its growth. The high fraction of CSS sources with a
brighter-when-closer behaviour suggests that jet-cloud interactions
are not so unlikely and it may cause an underestimate of the source
age. \\

\section{Concluding remarks}

It has been more than two decades that compact steep spectrum and
GHz-peaked spectrum radio sources are under investigation. During this
time some questions got an answer, perhaps not unique, and new
questions raised.\\
It is now fairly recognized that the majority of CSS and GPS radio
sources, at least
those optically associated with galaxy
represent an early stage in the radio source evolution. These objects
show two-sided structures similar to FRII radio galaxies, but on much
smaller scales. No significant variability is observed. However, some
level of variability explained in terms of adiabatic expansion in very
young objects have been observed in a few radio sources.\\
The distinctive characteristic of CSS/GPS objects, i.e. the presence of a
spectral turnover at 
low frequencies, is mainly due to SSA, although an additional
contribution from FFA is detected in the most compact objects. 
The
ambient medium enshrouding the radio emission is rich and inhomogeneous,
favouring interactions between the jet and dense clouds. The high
fraction of asymmetric sources, that is much higher than in extended
FRII galaxies, points out how frequently a cloud can impact with one
jet, temporarily confining, or at least slowing down, its advance
speed, and enhancing the synchrotron emission. This may explain part
of the excess in the number counts in flux limited sample. In fact,
CSS/GPS are too many, even if we assume a decrease of the luminosity
as they grow. This is predicted by the evolutionary models developed
taking into account the ambient medium, and is confirmed by
observations of samples of CSS/GPS spanning a large range of linear
size. \\
Following the evolutionary scenario, GPS radio sources are the
progenitors of CSS objects, which then should evolve in FRII radio
galaxies. However, it seems that some CSS/GPS are short-lived
objects, with ages up to 10$^{3}$ -- 10$^{4}$ years, which may never become
large sources with size of hundred of kpc or more. Only a handful of
CSS/GPS have been recognized to be in a relic phase on the basis of
the absence of active regions. These objects are under-represented in
flux-limited catalogues due to the steepness of their synchrotron
spectrum. The advent of the new radio facilities, like LOFAR, Merkaat,
ASKAP, and SKA, will provide a step forward in estimating the
incidence of short-lived and/or recurrent objects, providing a further
piece in the puzzle of our understanding of the radio emission
phenomenon.\\


\acknowledgements
I would like to thank Daniele Dallacasa for all the helpful and
constructive discussions about the topics presented in this work.\\

\end{document}